GRAVITATION,
ASTROPHYSICS

# The Formation of Primary Galactic Nuclei during Phase Transitions in the Early Universe

S. G. Rubin[a, b, *], A. S. Sakharov[c, **], and M. Yu. Khlopov[a, b, d, ***]

[a]*Moscow Engineering Physics Institute, Moscow, 115409 Russia*
[b]*Center for Cosmoparticle Physics "Cosmion", Moscow, 125047 Russia*
[c]*Labor für Höchenergiephysik, ETH-Hönggerberg, HPK-Gebäude, CH-8093, Zurich*
[d]*Institute for Applied Mathematics, Russian Academy of Sciences, Moscow, 125047 Russia*
*e-mail: serg.rubin@mtu-net.ru
**e-mail: sakhas@particle.phys.ethz.ch
***e-mail: mkhlopov@orc.ru
Received February 8, 2001

**Abstract**—A new mechanism describing the formation of protogalaxies is proposed, which is based on the second-order phase transition in the inflation stage and the domain wall formation upon the end of inflation. The presence of closed domain walls with the size markedly exceeding the cosmological horizon at the instant of their formation and the wall collapse in the postinflation epoch (when the wall size becomes comparable with the cosmological horizon), which leads to the formation of massive black hole clusters that can serve as nuclei for the future galaxies. The black hole mass distributions obtained do not contradict the available experimental data. The number of black holes with $M \sim 100$ solar masses ($M_\odot$) and above is comparable with the number of Galaxies in the visible Universe. Development of the proposed approach gives ground for a principally new scenario of the galaxy formation in the model of hot Universe. © *2001 MAIK "Nauka/Interperiodica"*.

## 1. INTRODUCTION

In the past decade, investigations into the nature of active galactic nuclei exhibited a considerable progress. Now there is virtually no doubt that the centers of galaxies contain massive black holes [1]. It is the existence of black holes with masses on the order of $10^6$–$10^8$ $M_\odot$ in the galactic nuclei and the accretion of matter onto these holes that is believed to account for the physical nature of their activity. A possible explanation of the formation of such supermassive black holes assumes the collapse of a large number of stars caused by their high concentration at the galaxy center. However, the mechanism of the galactic nuclei formation is still unclear. According to Veilleux [2], there are serious grounds to believe that the formation of stars and galaxies proceeded simultaneously. Stiavelli [3] considered a model of the galaxy formation around a massive black hole and presented arguments in favor of this model (see also [4]). Each of the two approaches has certain advantages, while being not free of drawbacks.

The problem of the possible "primordial" black hole (PBH) formation is still open. In contrast to the case of "secondary" black holes, which are related to the evolution of stars and stellar systems, there is no convincing astronomic evidence for the existence of PBHs. Restrictions posed by the astronomic data on the PBH concentration offer a unique source of information on the inhomogeneity of early Universe [5] and on the physical processes accounting for this inhomogeneity [6]. Generally speaking, the PBH mass may be arbitrary, ranging from the Planck value (or even below [7]) up to a mass contained within the contemporary cosmological horizon. However, in most cases the astrophysical effects related to the presence of PBH are restricted to masses much lower than the solar mass. The reason is that the mechanism of PBH formation is usually related to the development of inhomogeneities bounded by the cosmological horizon. The data of observations concerning the distribution of light elements and the spectrum of cosmic microwave radiation pose very rigid restrictions on the magnitude of inhomogeneities existing in the pregalactic stage following the first second of expansion of the Universe. Thus, realistic mechanisms of the PBH formation have to be apparently related to very early ($t \ll 1$ s) stages of evolution of the Universe—when the mass contained within the cosmological horizon and limiting the possible PBH mass was significantly lower than the solar mass. Nevertheless, the actively discussed possibility of a genetic relationship between quasars and active galactic nuclei on the one hand and the existence of PBHs with much greater masses on the other hand [8] becomes a subject for detailed investigations [9–11].

Below we will consider a new mechanism describing the early formation of PBHs, which serve as the nucleation centers in the subsequent formation of galaxies. This mechanism may prove to be free from dis-advantages inherent in the models based on the concept of a single PBH being a nucleus of the future galaxy.

Previously [11] we proposed a new mechanism of the PBH formation that opens possibility of the massive black hole formation in the early Universe. The mechanism is based on the possibility that black holes are formed as a result of a collapse of closed walls formed during a second-order phase transition. The masses of such black holes may vary within broad limits, up to a level on the order of $10^8 M_\odot$.

Let us assume that a potential of the field system possesses at least two different vacuum states. Then there are two possible distributions of these states in the early Universe. The first possibility is that the Universe contains approximately equal numbers of both states, which is typical of a temperature-controlled phase transition under usual conditions. The alternative possibility corresponds to the case when the two vacuum states form with different probabilities. In this case, there appear islands of less probable vacuum state surrounded by the sea of another, more probable vacuum state. As was recently demonstrated [12], an important condition for this distribution is the existence of valleys in the scalar field potential during inflation. Then the background de Sitter fluctuations in this massless scalar field lead to the formation of islands representing one vacuum in the sea of another vacuum. this phase transition takes place after the end of inflation in the Friedmann–Robertson–Walker Universe. After the phase transition, the two vacuum states are separated by a wall; the size of this wall may be significantly greater as compared to the cosmological horizon at that period of time. At some instant after crossing the horizon, the walls begin to contract because of the surface tension. As a result, provided that friction is absent and the wall does not radiate a considerable part of its energy in the form of scalar waves, almost all energy of this closed wall may be concentrated within a small volume inside the gravitational radius. This is a necessary condition of the black hole formation.

The mass spectrum of black holes formed by this mechanism depends on parameters of the scalar field potential determining the direction and size of the potential valley during inflation and the postinflation phase transition. Although we deal here with the so-called pseudo-Nambu–Goldstone field, the proposed mechanism possesses a sufficiently general character.

The presence of massive PBHs is a new factor in the development of gravitational instability in the surrounding matter and may serve a base for new scenarios of the formation and evolution of galaxies.

## 2. THE FORMATION OF CLOSED WALLS IN A COMPLEX FIELD

Now we will describe a mechanism accounting for the appearance of massive walls with the size markedly greater than the horizon at the end of inflation. Let us consider a complex scalar field with the potential

$$V(\varphi) = \lambda(|\varphi|^2 - f^2/2)^2, \quad (1)$$

where $\varphi = re^{i\theta}$. We assume the mass of the radial field component $r$ to be sufficiently large, so that the complex field would occur in the ground state even before the end of inflation. Since the minimum of potential (1) is degenerate, the field has the form

$$\varphi \approx (f/\sqrt{2})e^{i\theta},$$

with the phase $\theta$ acquiring the meaning of a massless field.

For the following considerations, it should be noted that, using expression (1) in the inflation period, we ignored the term

$$\delta V(\theta) = \Lambda^4(1 - \cos\theta). \quad (2)$$

reflecting the contribution of instanton effects to the Lagrangian renormalization. Since the parameter $\Lambda$ appears as a result of the normalization, its value cannot be large and we may quite reasonably assume that $\Lambda \ll H, f$. The omitted term (2) begins to play a significant role in the postinflation stage, when the Hubble constant sharply decreases with time ($H = 1/2t$ during the radiation dominated epoch).

Let us assume that a part of the Universe occurring inside the contemporary horizon was formed $N_U$ *e*-folds before the end of inflation. As was demonstrated in [13], the quantum field fluctuations during inflation are rapidly transformed into a classical field component, while the massless field values in the neighboring causality-disconnected intervals with the size $H^{-1}$ differ on the average by

$$\delta\theta = H/2\pi f \quad (3)$$

after a single *e*-fold. In the next time step $\Delta t = H^{-1}$ (i.e., during the next *e*-fold) each causality-connected domain is divided into $\sim e^3$ causality-disconnected subdomains; the phase in each of the new domains differs by $\sim \delta\theta$ from that in the preceding step. Thus, more and more domains appear with time in which the phase differs significantly from the initial value. More precisely, the probability of finding the phase $\theta$ is [14–16]

$$P(\theta, N) = \frac{1}{\sqrt{2\pi}\sigma_N}\exp\left\{-\frac{\theta_U - \theta}{2\sigma_N^2}\right\},$$

$$\sigma_N = \frac{H}{2\pi f}\sqrt{N_U - N}. \quad (4)$$

where $N$ is the number of *e*-folds remaining to the end of the inflation period, $\theta_U$ is the random phase value at the instant of formation of the causality-connected domain corresponding to a visible part of the contemporary Universe. Without a loss of generality, we may select $0 < \theta_U < \pi$. Below we will demonstrate that a par-ticular value of the initial phase significantly affects evolution of the Universe in the postinflation epoch.

The probability of finding a certain phase obeys the Gaussian distribution (4) and, hence, the phase averaged over the entire space equals the random initial phase $\theta_U$. A principally important point is the appearance of domains with the phases $\theta > \pi$. Appearing only after a certain period of time during which the Universe exhibited exponential expansion, these domains turn out to be surrounded by a space with the phase $\theta < \pi$. These very domains lead in the following to the formation of large-scale structures. Note that the phase fluctuations during the first $e$-folds may, generally speaking, transform eventually into fluctuations of the cosmic microwave radiation, which will lead to imposing restrictions on the scaling parameter $f$. This difficulty can be avoided by taking into account interaction of the field $\varphi$ with the inflaton field (i.e., by making parameter $f$ a variable).

Initially, the potential (1) possessed a $U(1)$ symmetry and the phase $\theta$ corresponded to a massless scalar field. Owing to the term (2), the symmetry is broken after the end of the inflation period: the potential of the $\theta$ field acquires minima at the points $\theta_{min} = 0, \pm 2\pi, \pm 4\pi, \ldots$, and the field acquires the mass $m_\theta = 2f/\Lambda^2$. According to the classical equation of motion,

$$\ddot{\theta} + 3H\dot{\theta} + \frac{d\delta V}{d\theta} = 0, \quad (5)$$

the phase performs decaying oscillations about the potential minimum, the initial values being different in various space domains. Moreover, domains with the initial phase $\pi < \theta < 2\pi$ perform oscillations about the potential minimum at $\theta_{min} = 2\pi$, whereas the phase in the surrounding space tends to a minimum at the point $\theta_{min} = 0$. Upon ceasing of the decaying phase oscillations, the system contains domains characterized by the phase $\theta_{min} = 2\pi$ surrounded by the space with $\theta_{min} = 0$. Apparently, on moving in any direction from inside to outside of the domain, we will unavoidably pass through a point where $\theta = \pi$ because the phase varies continuously. This implies that a closed surface must exist which is characterized by the phase $\theta_{wall} = \pi$. The size of this surface depends on the moment of domain formation in the inflation period, while the shape of the surface may be arbitrary. A principal point for the subsequent considerations is that the surface is closed.

Thus, we obtained a field configuration connecting various vacuum states of the potential (2). A rigorous classical solution of this problem possessing a translational symmetry in the two directions in space (flat wall) is well known [17]:

$$\theta_{wall}(x - x_0) = -4\arctan\left[\exp\left(\frac{x - x_0}{d}\right)\right], \quad (6)$$

where $d$ is the wall thickness. Since the thickness of a closed wall is related to microscopic parameters of the theory, whereas the characteristic wall size is *a priori* unlimited, expression (6) is applicable to within a sufficiently high accuracy in the case under consideration. As can be readily shown, the wall possesses an energy concentrated where the phase is $\theta = \pi$ [17]. Thus, we obtained a mechanism providing the formation of domains surrounded by closed walls. The surface energy of a wall depends on the Lagrangian parameters, while the wall size id determined by the time of crossing the phase value equal to $\pi$ during the inflation period.

## 3. DOMAIN WALL DECELERATION DURING MOTION THROUGH A PLASMA

The first- and second-order phase transitions lead to the formation of a field walls separating one vacuum of this field from another. One of these mechanisms was described in the preceding section. In turn, the walls are moving at a subluminal velocity and interact with the surrounding plasma. Depending on the character of this interaction and the shape of the field potential, there are two possible situations. In the first case, the plasma particles pass through the wall, falling into a different vacuum and acquiring a certain mass. This situation corresponds to an electroweak phase interaction [18], whereby the corresponding Higgs field is responsible for a mechanism of the fermion mass production. In the opposite case, the particle mass does not change upon going from one to another vacuum (an example is offered by the case of interaction with an axion wall). In the former case, the interaction with the medium leads to a significant retardation of the domain wall, while in the latter case, the wall are virtually transparent for the medium provided that the parameters are given reasonable values.

All considerations are conveniently conducted in the resting wall frame. The probability of a particle scattering from the plane resting wall is

$$dw = dn(k)2\pi\delta(\varepsilon - \varepsilon')|M|^2 \frac{d^3 k'}{2\varepsilon V(2\pi)^3 \varepsilon'}. \quad (7)$$

where $dn(k)$ is the distribution of incident particles with respect to momentum, $M$ is the matrix element for the particle transition from a state with the energy $\varepsilon$ and momentum $k$ to the state with the energy $\varepsilon'$ and momentum $k'$ upon interaction with the potential $U = U(z)$ describing the plane wall. The pressure produced by incident particles upon the wall is related to the rate of their momentum transfer to the wall

$$p = \frac{1}{S}\int dw q_z, \quad q_z = k'_z - k_z, \quad (8)$$

where $S$ is the wall area. Let us select the Lagrangian of the particle–wall interaction in the following form representing a classical configuration of the complex field phase:

$$L_{int} = \kappa \partial_z \theta(z) J_z, \quad J_\mu = \overline{\psi}\gamma_\mu\psi, \quad \kappa = i/f. \quad (9)$$

Calculating a matrix element for the particle scattering from the wall with the transition from the initial momentum $k$ to final momentum $k'$

$$M = \langle k'|\int L_{\text{int}} d^4 x |k\rangle \qquad (10)$$

we obtain

$$|M|^2 = 8(4\pi)^6 \kappa^2 S \delta^{(2)}(\mathbf{q}_\parallel) k_z^2 \frac{1}{\cosh^2(k_z d\pi)}. \qquad (11)$$

In deriving formula (11), we took into account that the laws of the energy–momentum conservation lead to the following relationships

$$k_z' = \pm k_z; \quad q_\parallel \equiv k_\parallel' - k_\parallel = 0,$$

according to which a nonzero contribution to the pressure is only due to the reflected particles with $k_z' = -k_z$. Therefore, the pressure of incident particles upon the wall can be written as

$$q = \frac{4}{\pi^2} \kappa^2 \int \frac{k_z^2}{\cosh^2(\pi k_z d)} (k_z - k_z') \delta(\varepsilon - \varepsilon') \qquad (12)$$
$$\times \delta(k_\parallel - k_\parallel') \frac{dn(k)}{V} \frac{d^3 k'}{\varepsilon \varepsilon'}.$$

Let us determine a distribution of the incident particles with respect to the transverse momentum $dn(k)$. In the resting plasma frame,

$$dn_0(k_0) = C \exp\left\{-\frac{E_0(\mathbf{k}_0)}{T}\right\} \frac{d^3 k_0 V}{(2\pi)^3}. \qquad (13)$$

Here and below, the subscript 0 denotes quantities determined in the resting plasma frame. Assuming the plasma temperature $T$ to be significantly greater as compared to the fermion masses and normalizing it to the total particle density

$$n_{\text{tot}} \approx N(g*) T^3, \quad N(g*) \approx 5,$$

we obtain $T \approx 5/8\pi$. In addition, it is evident that

$$dn(k) = dn_0(k_0), \qquad (14)$$

where the incident particle momentum in the resting wall and plasma frames (in the latter frame, the wall moves at a velocity $v$) are related as

$$k_{0\parallel} = k_\parallel,$$
$$k_{0z} = \gamma(k_z + v\varepsilon),$$
$$E_0 = \gamma(v k_z + \varepsilon), \qquad (15)$$
$$\gamma = \frac{1}{\sqrt{1-v^2}}.$$

Integrating the pressure (12) with resect to momentum of the incident particle, we obtain

$$p = \text{sgn}(k_z) \frac{32C\kappa^2}{(2\pi)^5} \gamma \int \frac{d^3 k}{\varepsilon^2} (\varepsilon + v k_z) \frac{k_z^2}{\cosh^2(\pi k_z d)}$$
$$\times \exp\left\{-\gamma \frac{\varepsilon + v k_z}{T}\right\}. \qquad (16)$$

This formula was derived with an allowance of the Lorentz invariance of the phase volume $d^3 k'/\varepsilon'$. Numerical calculation of the integral in (16) presents no difficulties, but we are interested in analytically estimating the pressure produced by the medium upon the wall. For this purpose, note that the walls are formed at temperatures $T \sim \Lambda$ and the wall thickness is $d = f/2\Lambda^2$. Therefore, there is a large parameter

$$Td \approx f/\Lambda \gg 1,$$

using which we may obtain a sufficiently reliable estimate of the integral. According to (16), the most effective scattering takes place for an incident particle momentum of

$$k_z \sim 1/\pi d \ll T.$$

At the same time, it is evident that

$$k_\parallel \sim \varepsilon \sim \gamma T \gg k_z.$$

Using these relationships, we may estimate the integral in (16). A final expression for the pressure produced by the surrounding medium upon the domain wall is as follows:

$$p \approx \frac{20\kappa^2}{\pi^7} \frac{\gamma}{d^4}. \qquad (17)$$

## 4. CONDITIONS FOR THE PRIMORDIAL BLACK HOLE FORMATION

After heating of the Universe, the evolution of domains formed with the phase $\theta > \pi$ and sharply increased in volume during the inflation period proceeds on the background of the Friedmann expansion and is described by the relativistic equation of state. First, an equilibrium state with the "vacuum" phase $\theta = 2\pi$ inside the domain and the $\theta = 0$ phase outside is established at $T \sim \Lambda$. A closed wall corresponding to the phase $\theta = \pi$ is formed in the transition region with a width of $\sim 1/m \sim f/\Lambda^2$, which separates the domain from the surrounding space. The surface energy density on the wall amounts to $\sim f/\Lambda^2$.

It must be noted that the process of establishing of the equilibrium ("vacuum") phase values may acquire a protracted character. If the stage of coherent phase oscillations about the equilibrium values is sufficiently long, the energy density of these oscillations may become dominating and determine the dust period of expansion. Let us consider the factors influencing the cosmological evolution of such a wall.

1. First, note that the domain size immediately after the end of inflation markedly exceeds the horizon size in the Friedmann expansion stage. The overall contraction of the closed wall may begin only when the horizon size $R_h$ will be equal to the domain size. Up to this moment, the characteristic domain size increases with the expanding Universe because we assumed that the existing field contribution to the total energy–momentum tensor is small as compared to that of the inflaton field. Accordingly, the field gives also a small contribution to the total energy density of the Universe upon heating, when the energy density of inflaton transforms into the energy density of relativistic particles. Evidently, internal stresses developed in the wall after crossing the horizon initiate processes tending to minimize the wall surface. This implies that the wall tends, first, to acquire a spherical shape and, second, to contract toward the center. For simplicity, below we will consider the motion of closed spherical walls.

2. Since the energy of the surrounding plasma rapidly decreases, the wall energy may become at a certain time instant comparable with the energy of the surrounding medium. Simultaneously, the domain separates from the general expansion process and its radius $R_w$ may become smaller than $R_h$.

3. The wall energy is proportional to its area at the instant of crossing the horizon. By the moment of maximum contraction, this energy is virtually completely converted into the kinetic energy. Should the wall by this moment be localized within the limits of the gravitational radius, a PBH is formed.

4. Contracting under the action of internal forces, the wall moves through the surrounding plasma. The resulting force of friction may, under certain conditions, become significant and lead to a uniform (nonaccelerated) contraction of the wall. In this case, the potential energy of the wall is dissipated in the surrounding medium. Only when the wall would decrease to a certain small size $R_f$, the internal forces proportional to the surface curvature will dominate and the wall will again contract with acceleration to supply a necessary energy to the center, sufficient to form a PBH.

The above considerations show that the energy concentrated in the course of wall contraction can be determined using the condition

$$E \approx 4\pi R^2 \sigma, \quad R = \min(R_h, R_w, R_f). \quad (18)$$

where $\sigma = 4\Lambda^2 f$ is the surface energy density of the wall. A condition of the PBH formation is that

$$R_{\min} \sim d < r_g = 2E/m_{pl}^2. \quad (19)$$

It is assumed that the spherical wall contracts until reaching a size on the order of the wall thickness.

Let us determine the values of $R_h$, $R_w$, and $R_f$ for a system with Lagrangian (1). Consider a domain with a certain phase appearing $N$ $e$-folds before the end of inflation. The domain size at the end of inflation period is

$$l_e = H^{-1} e^N, \quad (20)$$

where $H$ is the Hubble constant at that time instant. By the moment of crossing the horizon, the domain will expand to acquire the characteristic size

$$R_h \approx \frac{e^{2N}}{2HN}. \quad (21)$$

Below we assume that a visible part of the Universe is formed $N = N_U = 60$ $e$-folds before the end of inflation.

The second characteristic size $R_w$ is determined from the condition that the wall energy (18) is equal to the energy of plasma contained in the domain bounded by the closed wall:

$$E_V = \rho \frac{4\pi}{3} R^3$$

where

$$\rho = \frac{\pi^2}{30} g^* T^4$$

is the plasma density during the radiation dominated epoch. Taking into account that $\sigma = 4\Lambda^2/f$, we obtain the critical wall size corresponding to the domain separating from the general expansion:

$$R = R_{\text{crit}} = \frac{3\sigma}{\rho} = \frac{360\Lambda^2 f}{\pi^2 g^* T^4} \approx \frac{\Lambda^2 f}{T^4}. \quad (22)$$

As is known, the temperature in the Robertson–Walker Universe during the radiation dominated epoch varies with time as

$$T = \left(\frac{45}{32\pi^2 g^*}\right)^{1/4} \sqrt{\frac{m_{pl}}{t}}. \quad (23)$$

Taking into account that an increase in the wall radius up to the instant of separation from the general expansion is proportional to the scaling factor

$$R(t) = l_e \sqrt{t/t_e}, \quad (24)$$

we arrive at the desired relationship for $R_w(N)$:

$$R_w = (2^8 \pi \sigma)^{-1/3} \left(\frac{m_{pl}}{H}\right)^{2/3} N^{-2/3} \exp\left(\frac{4N}{3}\right). \quad (25)$$

An expression for the characteristic wall size (radius) $R_f$ above which the friction is significant can be obtained by equating the pressure developed by the internal forces $p_{\text{int}} = 2\sigma/R_f$ to that produced by the surrounding medium on the moving wall. Using relationship (17) for the latter pressure, we obtain

$$R_f = \frac{\pi^7 \sigma d^4}{10 \kappa^2 \gamma}. \quad (26)$$

The above considerations do not take into account the effect of a gravity field on the wall dynamics. Therefore, the obtained relationships are valid provided that the initial wall size is much greater than the gravitational radius. Generally speaking, the gravitational radius may be comparable with (or even exceed) the wall size for a sufficiently large domain size. In this study, we deal with smaller domains for which the intrinsic gravity field does not affect the wall evolution. Now we proceed to the study of PBH cluster formation in the early evolution stage of the Universe.

## 5. CORRELATIONS IN THE BLACK HOLE DISTRIBUTION

Previously [11] we have studied a new process involving the formation of uncorrelated PBHs in the Universe. It was demonstrated that a model with reasonably selected parameters readily provides for the formation of $10^{11}$ massive ($10^{30}$–$10^{40}$ g each) black holes, which is equal precisely equal to the number of Galaxies in the visible Universe. In that analysis, we did not took into account correlations (inherent in this mechanism) between the formation of a massive black hole and the appearance of smaller back holes surrounding the former one. This correlation is related primarily to certain features of the above-discussed process of the formation of domains with the phases $\theta > \pi$. Apparently, the appearance of such domains creates prerequisites for the formation of new smaller domains inside.

Let us estimate the mass distribution of these daughter domains. Consider a region with a size on the order of $H^{-1}$ and a phase within $\pi < \theta_0 < \pi + \delta$ (where $\delta = H/2\pi f$ is the average phase jump during the $H^{-1}$ time period) formed during the inflation period as a result of fluctuation in a certain region of space with the phase $\theta < \pi$. During the next $e$-fold, this space domain will separate into $e^3$ subdomains and some of these will acquire a phase $\theta_1$ in the interval $\pi - \delta < \theta_1 < \pi$. Upon the subsequent phase transition, these domains will be separated by walls from the external region. Similar transitions, with crossing the phase $\theta = \pi$ in the reverse direction will take place in each subdomain during the next $e$-fold. Thus, a structure of the fractal type appears which reproduces itself in each time step on a decreasing scale.

Let $\zeta$ denote the number of subdomains formed in each step, around which a wall may form with time. Apparently, this value obeys the inequality $1 < \zeta \ll e^3$. In the subsequent estimates, we will assume that $\zeta \approx 2$–$3$. Since each causality-connected domain touches approximately six neighboring domains, we can hardly expect $\zeta$ to be greater for a total number of $\sim e^3 \approx 20$. The mass of the future black hole (if this would actually form) is determined by the area of a closed surface with the phase $\theta = \pi$. The ratio of areas of the initial (mother) and daughter domains is readily estimated: the initial area after a single e-fold is

$$S_0 \approx e^2 H^{-2},$$

and the daughter subdomain area is

$$S_1 \approx H^{-2}.$$

Therefore, the ratio of masses of the black holes belonging to two sequential generations is

$$M_j / M_{j+1} \approx S_j / S_{j+1} \approx e^2, \quad (27)$$

for their relative number assumed to be

$$N_{j+1} / N_j = \zeta. \quad (28)$$

As is readily seen, the number and mass of black holes appearing upon the $j$th $e$-fold after the initial domain formation are determined by parameters of the largest black hole genetically related to the primary domain in which the phase originally exceeded $\pi$. It is evident that

$$N_j \approx \zeta^j, \quad M_j \approx M_0 / e^{2j}. \quad (29)$$

Excluding $j$ from these relationships, we obtain the desired black hole mass distribution in a cluster:

$$N_{cl}(M) \approx (M_0/M)^{(1/2)\ln\zeta}. \quad (30)$$

The total mass of the cluster can be expressed through the mass $M_0$ of the largest initial black hole:

$$M_{tot} = M_0 + \zeta M_1 + \zeta^2 M_2 + \ldots = M_0 \\ + \zeta e^{-2} M_0 + (\zeta e^{-2})^2 M_0 + \ldots = M_0 [1 - \zeta/e^2]^{-1}. \quad (31)$$

As is seen, the total mass of the black hole cluster is only one and a half to two times greater than the largest initial black hole mass. The number of daughter black holes depends on the factors considered in the next section.

## 6. DISCUSSION OF RESULTS

In the preceding sections, we considered only the principal possibility of the formation of domain walls connecting adjacent vacuum states. We have used the formulas derived above to estimate efficiency of the proposed mechanism of the black hole cluster formation. The numerical calculations were performed for the following values of parameters (which are consistent with the observed anisotropy in the cosmic microwave radiation): the Hubble constant at the end of inflation, $H = 10^{13}$ GeV; Lagrangian parameters, $f = 1.77H$ and $\Lambda = 5$ Gev. The initial phase, at which the visible part of the Universe is formed by the time $t_U \approx 60 H^{-1}$ to the end of inflation, controls the number of domains and, accordingly, the number of closed walls formed in the postinflation stage. This random value, not related to the Lagrangian parameters, must be selected taking into account the data of observations on the abundance of black holes in the Universe. We will use the numerical value $\theta_U = 0.05\pi$, which ensures a sufficiently large number of massive black holes, while the presence of numerous smaller black holes does not contradict experimental restrictions.

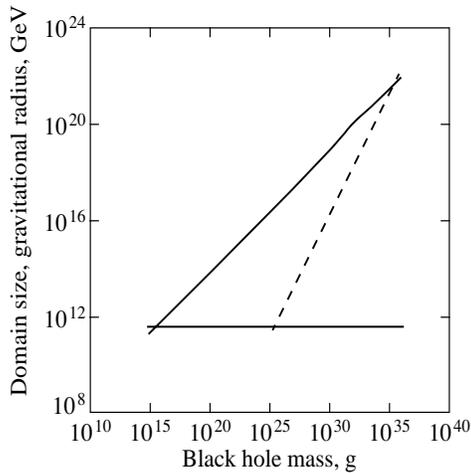

**Fig. 1.** Plots of the characteristic domain size (inclined solid line) and gravitational radius (dashed line) versus the domain mass. Horizontal line indicates the wall thickness. See the text for numerical values of parameters.

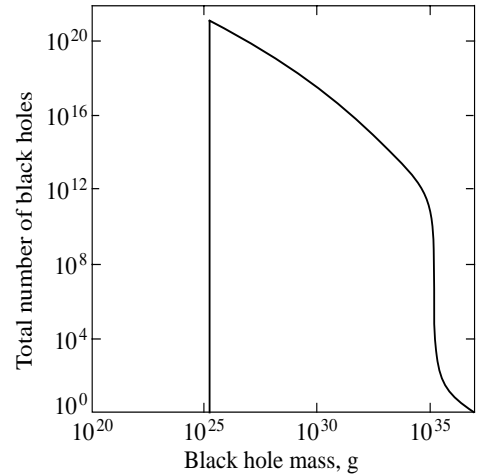

**Fig. 2.** Primordial black hole mass distribution in the Universe.

Figure 1 shows the results of numerical calculations constructed in the logarithmic scale. The bottom horizontal line shows the wall thickness. As is seen, the condition of wall existence (the characteristic domain size must exceed the wall thickness) is fulfilled for the domains with masses exceeding $10^{15}$ g. The domains of lower energies possess (for the parameters selected) dimensions below the wall thickness. This implies that the wall formation is impossible and the domain exhibits only fluctuations in the energy density. During contraction, the wall energy is eventually completely converted into radiation.

The necessary condition for a black hole formation as a result of the domain wall collapse is that the gravitational radius of the domain must be greater than the wall thickness. As is seen from fig. 1, this condition is fulfilled for the black holes with masses $\geq 10^{25}$ g. Therefore, the proposed mechanism of the black hole formation leads to a nontrivial situation: massive PBHs exist at a compete absence of the black holes possessing masses below this threshold. Note that most significant observational restrictions concerning the PBH abundance refer to the mass region $\sim 10^{15}$ g (which cannot form for the parameters selected). Since the gravitational radius is proportional to the wall surface area, the plots corresponding to the domain size and its gravitational radius must intersect. This intersection actually takes pace at a wall mass of $10^{35}$ g. For greater masses, the gravity effects have to be taken from the very beginning, which will limit from above the possible PBH masses. According to formula (26), the friction becomes significant only for supermassive walls, the number of which is negligibly small.

Figure 2 shows the PBH mass distribution calculated for the selected parameters (see also the discussion in [11]). As is seen, the PBH masses fall within the range from $10^{25}$ to $10^{35}$ g. The initial phase $\theta_U$ was selected so as to provide that the number of massive PBHs ($\sim 10^{35}$ g) was equal to the number of Galaxies in the visible part of the Universe. The total mass of black holes amounts to ~1% of the contemporary baryonic contribution.

The results of calculations are sensitive to changes in the parameter $\Lambda$ and the initial phase $\theta_U$. As the $\Lambda$ value decreases to $\approx 1$ GeV, still greater PBHs appear with a mass of up to $\sim 10^{40}$ g. A change in the initial phase leads to sharp variations in the total number of black holes. As was shown in Section 5, each domain generates a family of subdomains in the close vicinity. A total mass of such a cluster is only 1.5–2 times that of the largest initial black hole in this space region. Thus, our calculations confirm the possibility of formation of the clusters of massive PBHs ($\sim 100$ $M_\odot$ and above) in the earliest stages of evolution of the Universe at a temperature of 1–10 GeV. These clusters represent stable energy density fluctuations around which increased baryonic density may concentrate in the subsequent stages, followed by the evolution into Galaxies.

## 7. CONCLUSION

This paper proposes a new mechanism for the formation of protogalaxies, which is based on the cosmological inferences of the elementary particle models predicting nonequilibrium second-order phase transition in the inflation stage period and the domain wall formation upon the end of inflation. The presence of closed domain walls with the size markedly exceeding the cosmological horizon at the instant of their formation leads to the wall collapse in the postinflation epoch (when the wall size becomes comparable with the cosmological horizon), which results in the formation of massive black hole clusters that can serve as nuclei for the future galaxies. The black hole mass distributions are calculated which do not contradict the available experimental data. The number of black holes with $M \sim 100\, M_\odot$ and above is comparable with the number of Galaxies in the visible Universe. A mechanism of deceleration of the wall motion is considered and it is shown that this process may affect only the dynamics of collapse of supermassive walls.

Development of the proposed approach gives ground for a principally new scenario of the galaxy formation in the model of hot Universe. Traditionally, the hot Universe model assumes a homogeneous distribution of matter on all scales, whereas the appearance of observed inhomogeneities is related to the growth of small initial density perturbations. However, an analysis of the cosmological inferences of the theory of elementary particles indicates the possible existence of strongly inhomogeneous primordial structures in the distribution of both the latent mass and baryons. These primordial structures represent a new factor in theory of galaxy formation. Topological defects such as the cosmological walls and filaments, primordial black holes, archioles in the models of axion cold latent mass [19, 20], and essentially inhomogeneous baryosynthesis (leading to the formation of antimatter domains in the baryon-asymmetric Universe) [12, 21] offer by no means a complete list of possible primary inhomogeneities inferred from the existing elementary particle models.

The proposed approach discloses a number of interesting aspects in this direction. Indeed, this model provides for a possibility of the quantitative analysis of correlations in the formation of massive PBHs and the primary inhomogeneity of the latent mass and baryons. Originally inherent in this mechanism is the inhomogeneous phase distribution which eventually acquires (similar to what takes place in the invisible axion cosmology) a dynamical sense of the initial amplitude of the coherent oscillations of a scalar field. Irrespective of the efficiency of dissipation of the energy of these oscillations, the regions of closed wall formation must be correlated with the regions of maximum energy density of the latent mass. If these oscillations are not decaying, their energy density may provide for the contemporary latent mass density. Inhomogeneity in the initial amplitude of these oscillations would then imply an inhomogeneity in the initial energy density and, hence, the regions of black hole formation will become the regions of increased latent mass density. Qualitatively similar effect (albeit not as pronounced) takes place in the dissipation of coherent oscillations at the expense of particle production. An increase in the oscillation energy density transforms into a local increase in the density of latent mass particles produced in this region.

In the class of spontaneous baryosynthesis models, a change in the phase determines the production of excess baryons. Therefore, in addition to an increase in the latent mass density, the regions of massive PBH formation may be characterized by a higher baryon density. Inside a closed wall, where the phase is $\theta > \pi$, the same mechanism leads to the production of excess antibaryons [12]. However, this antimatter domain will survive only provided that its size is sufficiently large [22]. Thus, development of the proposed approach may lead to a number of interesting scenarios of initial stages in the formation of protogalaxies, depending on the selection of particular elementary particle models and their parameters. This study presents the first step in this direction.

## ACKNOWLEDGMENTS.

The authors are grateful to E.F. Zhizhin for the fruitful discussion or results and to A.A. Starobinsky for his interest in this study. The work of S.G.R. and M.Yu.Kh. was partly performed within the framework of the "Cosmoparticle Physics" Project of the State Scientific-Technological Program "Astronomy. Fundamental Space Research" supported by the Cosmion–ETHZ and Epcos–AMS international cooperation (Scientific School Support Grant no. 00-15-96699).

## REFERENCES

1. D. E. Rosenberg and J. R. Rutgers, astro-ph/0012023 (2000).
2. S. Veilleux, astro-ph/001212 (2000).
3. M. Stiavelli, astro-ph/9801021 (1998).
4. M. R. Merrifield, D. A. Forbes, and A. I. Terlevich, astro-ph/0002350 (2000).
5. I. D. Novikov, A. G. Polnarev, A. A. Starobinskiĭ, and Ya. B. Zel'dovich, Astron. Astrophys. **80**, 104 (1979).
6. A. G. Polnarev and M. Yu. Khlopov, Usp. Fiz. Nauk **145**, 369 (1985) [Sov. Phys. Usp. **18**, 213 (1985)].
7. Ya. B. Zel'dovich, Zh. Éksp. Teor. Fiz. **42**, 641 (1962) [Sov. Phys. JETP **15**, 446 (1962)].
8. M. Yu. Khlopov, Soobshch. Spets. Astrofiz. Obs. **6**, 7 (1987).
9. J. Yokoyama, astro-ph/9802357 (1998).
10. Hee Il Kim, Phys. Rev. D **62**, 063504 (2000); astro-ph/9907372 (1999).
11. S. G. Rubin, M. Yu. Khlopov, and A. S. Sakharov, Gravitation Cosmology, Suppl. **6**, 51 (2000).

12. M. Yu. Khlopov, S. G. Rubin, and A. S. Sakharov, Phys. Rev. D **62**, 083505 (2000).
13. A. A. Starobinskiĭ, Pis'ma Zh. Éksp. Teor. Fiz. **30**, 719 (1979) [JETP Lett. **30**, 682 (1979)].
14. A. Vilenkin and L. Ford, Phys. Rev. D **26**, 1231 (1982).
15. A. D. Linde, Phys. Lett. B **116**, 335 (1982).
16. A. Starobinsky, Phys. Lett. B **117**, 175 (1982).
17. R. Rajaraman, *Solitons and Instantons* (North-Holland, Amsterdam, 1982).
18. B. H. Liu, L. Mclerran, and N. Turok, Phys. Rev. D **46**, 2668 (1992).
19. A. S. Sakharov and M. Yu.Khlopov, Yad. Fiz. **57**, 514 (1994) [Phys. At. Nucl. **57**, 485 (1994)].
20. A. S. Sakharov, D. D. Sokolov, and M. Yu. Khlopov, Yad. Fiz. **59**, 1050 (1996) [Phys. At. Nucl. **59**, 1005 (1996)].
21. M. Yu. Khlopov, *Cosmoparticle Physics* (World Scientific, Singapore, 1999).
22. M. Yu. Khlopov *et al.*, Astropart. Phys. **12**, 367 (2000).

*Translated by P. Pozdeev*